\documentstyle[preprint,aps]{revtex}
\tightenlines

\begin{document}
\newcommand{\beq}{\begin{equation}}
\newcommand{\eeq}{\end{equation}}
\newcommand{\beqa}{\begin{eqnarray}}
\newcommand{\eeqa}{\end{eqnarray}}
\newcommand{\sr}{\sqrt}
\newcommand{\fr}{\frac}
\newcommand{\mn}{\mu \nu}
\newcommand{\G}{\Gamma}

\draft
\preprint{ INJE-TP-01-01, hep-th/0101091 }
\title{ Role of the brane curvature scalar in the brane world cosmology}
\author{  N. J. Kim\footnote{E-mail address:
dtpnjk@ijnc.inje.ac.kr},H. W. Lee\footnote{E-mail address:
hwlee@physics.inje.ac.kr}, and Y.S. Myung\footnote{E-mail address:
ysmyung@physics.inje.ac.kr} }
\address{
Department of Physics, Graduate School, Inje University,
Kimhae 621-749, Korea}
\maketitle

\begin{abstract}
We  include the brane curvature scalar to study its
 cosmological implication in the brane world cosmology.
 This term  is usually introduced to obtain the well-defined
 stress-energy tensor on the boundary of anti de Sitter-Schwarzschild  space.
 Here we treat this as the perturbed term for cosmological purpose.
 We find  corrections to the well-known equation of the brane
 cosmology.
 It contains new interesting terms which may play the important role in the
brane cosmology.
\end{abstract}

\newpage

Recently there has been much interest in the phenomenon of
localization of gravity proposed by Randall and Sundrum
(RS)~\cite{RS1,RS2}.
RS assumed
a single positive tension 3-brane and a negative bulk cosmological
constant in the 5D spacetime~\cite{RS2}. They have obtained a 4D localized gravity
by fine-tuning the tension of the brane to the cosmological constant.
The introduction of branes in a 5D anti de Sitter (AdS$_5$) space
usually gives rise to a non-compact extra dimension.

More recently, several authors have studied its cosmological
implications.
The brane cosmology   contains  some important
deviations from the Friedmann-Robertson-Walker(FRW) cosmology.
One  approach is first to assume the 5D dynamic metric (that is, BDL-metric\cite{BDL1,BDL2})
 which is
manifestly $Z_2$-symmetric, $\hat w \to -\hat w
$\footnote{ $\hat w$ is the fifth coordinate and the
metric is expressed in terms of the Gaussian normal coordinates.}.
Then one solves the Israel junction
condition \cite{ISR} and Einstein equation to find the behavior of the scale factor.
 We call its solution as the BDL
cosmological solution.

The other approach starts with
a static configuration which is two
5D anti de Sitter-Schwarzschild (AdSS$_5$) spaces joined by the domain wall.
 In this case  the
embedding into the moving domain wall is possible by choosing an appropriate
normal vector $n_M$\cite{CR,KRA,IDA}.
The domain wall separating two such bulk spaces is taken to be
located at $r=a(\tau)$, where $a(\tau)$ will determined by solving
the Israel junction condition. Then observers on the wall will
interpret their motion through the static bulk background as
cosmological expansion or contraction. Mukhoyama et al.\cite{MSM} performed
a coordinate transformation $\{\hat \tau,\hat x_i,\hat w \}$ $\to$
$\{t,r,\chi,\theta,\phi\}$ in order to bring the BDL metric into
the AdSS$_5$-metric. In this approach if two masses of
Schwarzschild black holes are different, one can easily find a situation
that does not possess a $Z_2$-symmetry manifestly
\footnote{In ref.\cite{DVD},
the authors derived a non-$Z_2$ symmetric term within the BDL approach. However,
it is hard to find the origin of non-$Z_2$ symmetric term in this approach.
Alternatives for non-$Z_2$ symmetric case in the brane physics, see ref.\cite{KLM}}.
We follow this approach.  This gives us
an asymmetric, cosmological evolution.

On the other hand, in the  AdSS$_5$-space, its boundary metric
acquires a divergent energy-momentum tensor $\Pi^{\mu\nu}$ as we take $r$ to infinity.
In order to have a well-defined asymptotically AdSS$_5$ space
($R\times S^3$) on the boundary near infinity,
 we need the counter terms to cancel the divergence
from the bulk\cite{BK}.
This is ${\cal L}_{ct}= -\fr{3}{\ell}\sqrt{-h}(1-
\fr{\ell^2}{12}{\cal R})$,
where  ${\cal R}$
 is the intrinsic curvature term on the boundary (domain
wall). In  view of the AdS/CFT correspondence, this inclusion of
counter terms might be interpreted as the expectation value of the energy-momentum
tensor in  quantum conformal field theory.

In this letter, we investigate the cosmological implication of the curvature scalar
on the brane. After the fine-tuning, this scalar can be of the
same order as the terms that remain in the field equations for
gravity on the brane\cite{CH}. Including this curvature scalar with
the perfect fluid as the localized matter on the brane, one finds
the fourth-oder equation for $k+ \dot a(\tau)^2$.
 Although the general solution to this equation is known, it is
 very hard to obtain a reliable equation like the FRW equation.
Hence  we consider the brane curvature scalar as the perturbation
term. Then we  find corrections to the known brane cosmology.
This can be considered as   one way to see  the effect of  brane
curvature scalar on the brane cosmology.

We will derive the Einstein equations on an 4D hypersurface
embedded in an 5D bulk space. We divide  the bulk space as two
regions, ${\cal M}_+$ and ${\cal M}_-$ separated by the domain
wall (brane), ${\cal B}$. Here "$+(-)$" denote  the right (left)-hand
sides. We  want to choose the different metric on both sides of the brane
but with the same cosmological constant and $k$ for
simplicity\footnote{ Two papers in \cite{STW,DD} have dealt with
the case of different cosmological constants. But this case does not give rise to
a non-$Z_2$ symmetric evolution at $a^{-6}$-order.}.
Also we include the  brane  curvature scalar ${\cal R}$ and
 the localized matter ${\cal L}_m$ including the brane tension as the
brane action. At each point on the brane, we introduce a normal
vector $n_M$ to the hypersurface such that $g^{MN}n_Mn_N=1$, where
$g_{MN}$ is the bulk metric and the capital indices $M,N,\cdots$ run over all bulk
coordinates. Then the induced metric on the brane  is given by
the tangential components of the projection tensor
\beq
h_{MN}=g_{MN}-n_{M}n_{N}
\label{INH}.
\eeq
We start by combining  two 5D bulk actions  and two Gibbons-Hawking
boundary terms\cite{CH}
\beqa
S_{1}=\fr{1}{16 \pi G} \int_{{\cal{M}}_+} d^5 x \sqrt{-g}
\left[R+ \fr{12}{\ell^{2}}
\right]+\fr{1}{8 \pi G} \int_{\cal{B}} d^4x \sqrt{-h}K^+,
\nonumber \\
S_{2}=\fr{1}{16 \pi G} \int_{{\cal{M}}_-} d^5 x \sqrt{-g}
\left[R+ \fr{12}{\ell^{2}}
\right]+\fr{1}{8 \pi G} \int_{\cal{B}} d^4x \sqrt{-h}K^-
\label{BAC}
\eeqa
and the brane action\footnote{ The cosmological application of
the brane curvature scalar ${\cal R}$ discussed
in refs.\cite{DEF,ANO}. Deffayet in \cite{DEF}
 considered this term to derive the Friedmann-like equation in 5D Minkowski spacetime, while
 authors in \cite{ANO} pointed out the existence of ${\cal R}$
 within the AdS/CFT correspondence.
}
\beq
S_b=\fr{1}{16 \pi G} \int_{\cal{B}} d^4 x \sqrt{-h}
\left[ b \fr{\ell}{2}{\cal R} +16 \pi G {\cal{L}}_m
\right],
\label{BRT}
\eeq
where the brane tension term is included at ${\cal L}_m=-\sigma + \cdots$.
The parameter $b$ is  introduced for our cosmological purpose. For a
counter term, we choose $\sigma=\fr{3}{8 \pi G \ell}, b=1$ \cite{BK}. In order
to obtain the RS Minkowski brane\cite{RS2},  one has to choose
$\sigma=\fr{6}{8 \pi G \ell}, b=0$.
But we regard $b$ here as an arbitrary, small parameter.
Here $G$ is the 5D Newtonian constant and the trace of the extrinsic curvature $K$ is
 introduced to obtain the well-defined variation on the brane (boundary).
The extrinsic curvature is defined by
\beq
K_{MN}={h_{M}}^{P} \nabla_{P} n_{N}.
\label{EXC}
\eeq
 The Einstein equation   for both sides leads to
\beq
G_{MN}=R_{MN}-\fr{1}{2}Rg^{\pm}_{MN}=\fr{12}{\ell^2}g^{\pm}_{MN},
\label{BEQ}
\eeq
where we choose the same cosmological constants on the two sides.
Further, from the boundary variation, one  takes the form
\beq
\triangle K_{MN}-h_{MN} \triangle K =b \fr{\ell}{2}
\left[{\cal{R}}_{MN} -\fr{1}{2}{\cal{R}}h_{MN}
\right]-8 \pi G T_{MN},
\label{BJC}
\eeq
where
\beq
\triangle K_{MN}\equiv K_{MN}^+-K_{MN}^-,~~~T_{MN}=
h_{MN}{\cal L}_m - 2 \fr{\delta {\cal L}_m}{\delta h^{MN}}.
\eeq
From Eq.(\ref{BJC}), we obtain the Israel junction condition
\beq
\triangle K_{MN}= -\kappa^2
\left[T_{MN}-\fr{1}{3}T_{P}^{P} h_{MN}
\right]
+ b \fr{\ell}{2}
\left[{\cal{R}}_{MN} -\fr{1}{6}{\cal{R}}h_{MN}
\right]
\label{IJC}
\eeq
with $\kappa^2= 8\pi G$.
This describes the relation of
the discontinuity of the extrinsic curvature across the surface
(brane)
to the surface stress-energy tensor, when the last term is absent.

Now we are in a position to discuss the static bulk solution to Eq.(\ref{BEQ})\cite{BIR}.
For the cosmological embedding, the relevant solution is chosen as the
AdSS$_5$-spacetime for both sides,
\beq
ds^{2}_{5\pm}=-h_\pm(r)dt^2 +\fr{1}{h_\pm(r)}dr^2 +r^2
\left[d\chi^2 +f_{k}(\chi)^2(d\theta^2+ \sin^2 \theta d\phi^2)
\right],~~~~(k=0,\pm1)
\label{BMT}
\eeq
where
\beq
h_\pm(r)=k-\fr{\alpha_\pm}{r^2}+ \fr{ r^2}{\ell^2},~~~
f_{0}(\chi) =\chi, ~f_{1}(\chi) =\sin \chi, ~f_{-1}(\chi) =\sinh
\chi.
\eeq
In the case of $\alpha_\pm=0$, we have two AdS$_5$-spaces.  However,
$\alpha_\pm \not=0$ generates the electric part of the Weyl tensor $C^\pm_{MNPQ}$
on each side. Its presence  means that the bulk
spacetime has the black hole horizon at small $r=r_{\pm h}= \ell^2(-k +\sqrt{k^2 +4
\alpha_\pm/\ell^2})/2$\cite{KRA}. On the other hand, we assume that the domain wall (brane)
is located at large $r$. For large $r$,
its asymptote to AdS$_5$-space is  $R \times S^3$. In this case,
the boundary energy-momentum tensor $\Pi^{\mu\nu}$ diverges as $r$
approaches infinity. It is due to the presence of $ r^2/\ell^2$-term in
$h_\pm(r)$. This corresponds to the intrinsic property of AdS$_5$.
Hence, to have a finite boundary energy-momentum tensor, one needs the  boundary
counter term such as ${\cal L}_{ct}= -\fr{3}{\ell}\sqrt{-h}(1-
\fr{\ell^2}{12}{\cal R})$.
However, the mass-term of $\alpha_\pm/r^2$ has the extrinsic
property from the boundary point of view. Because of its extrinsic property,
the $\alpha$-term cannot be subtracted off.
Hence we have to keep this  term for the brane cosmology.  We
expect that its interaction with the brane curvature scalar  plays an important
role.  Furthermore, if
$\alpha_+=\alpha_-$, we  find  a $Z_2$ symmetrical evolution of
the domain wall\cite{BDL2}. Different $\alpha$ ($\alpha_+ \not=\alpha_-$)
induces a non-$Z_2$ symmetric cosmology\cite{KRA,IDA,CH}.

Now we consider the location of brane (domain wall) in the form of
$t=t(\tau), r=a(\tau)$ parametrized by the proper time $\tau$ on
the brane: $(t,r,\chi,\theta,\phi)\to(t(\tau),a(\tau), \chi, \theta, \phi)$.
Then the induced metric of dynamical domain wall will be given
by the conventional FRW-type.
Here $\tau$ and $a(\tau)$ mean the cosmic time and scale factor of
the FRW universe, respectively. The tangent vectors of this brane
can be expressed as
\beq
u_\pm= \dot t_\pm \fr{\partial}{\partial t_\pm}+ \dot a \fr{\partial}{\partial
a},
\eeq
where the overdot means the differentiation with respect to
$\tau$. These satisfy $u_{\pm M}u_{\pm N} g^{MN}=-1$.
To find a dynamical solution, we need the normal 1-forms
 directed toward to each side: $n_{\pm M}n_{\pm N} g^{MN}=1$. Here we choose these as
\beq
n_\pm=\pm \dot a dt_\pm \mp \dot t_\pm da.
\eeq
This case is consistent with the Randall-Sundrum case in the limit
of
$\alpha_\pm=0$\cite{RS2}. Using this, we can express $\dot t$ in terms of $\dot a$
as
\beq
\dot t_\pm=\fr{(\dot a^2 +h_\pm(a))^{1/2}}{h_\pm(a)}.
\label{TAV}
\eeq
From the bulk metric Eq.(\ref{BMT}) together with Eq.(\ref{TAV}), we can derive the
4D induced metric
\beqa
ds^{2}_{4}&&=-d \tau^2 +a(\tau)^2
\left[d\chi^2 +f_{k}(\chi)^2(d\theta^2+ \sin^2 \theta d\phi^2)
\right]
\nonumber \\
&&\equiv h_{\mu \nu}dx^{\mu} dx^{\nu}.
\label{INM}
\eeqa
Hereafter we use the Greek indices  for  physics of the brane.
The extrinsic curvatures  for cosmological purpose are given by

\beqa
&&(K_\pm)_{\tau\tau}=(K_\pm)_{MN} u^M_\pm u^N_\pm =\pm(h_\pm \dot t_\pm)^{-1}(\ddot a +h'_\pm /2), \\
&&(K_\pm)_{\chi}^{\chi} = (K_\pm)_{\theta}^{\theta}=(K_\pm)_{\phi}^{\phi}
=\mp h_\pm \dot t_\pm /a,
\eeqa
where the prime stands for the derivative with respect to $a$.
The above equation implies that the extrinsic curvature jumps  across the brane.
This jump is already realized through the Israel condition of
Eq.(\ref{IJC}).  Here we need its four-dimensional version
\beq
\triangle K_{\mu \nu}=-\kappa^2 \left(
T_{\mu\nu}-\fr{1}{3}T^{\lambda}_{\lambda}h_{\mu\nu} \right)
+\fr{b\ell}{2}
\left(
{\cal R}_{\mu\nu}-\fr{1}{6}{\cal R}h_{\mu\nu} \right).
\label{4DI}
\eeq
In order to have a cosmological solution,
let us choose the localized stress-energy tensor on the
brane as the 4D perfect fluid

\beq
T_{\mu \nu}=(\varrho +p)u_{\mu}u_{\nu}+p\:h_{\mu\nu},
\label{MAT}
\eeq
Here $\varrho=\rho+ \sigma(p=P-\sigma)$, where $\rho(P)$ is the energy density (pressure)
of the localized matter and $\sigma$ is the brane tension.
In the absence of a localized matter, the first term of
Eq.(\ref{4DI}) takes the form of the RS case as $-\fr{\sigma \kappa^2}{3}
h_{\mu\nu}$.
 We stress again that  $u_\mu,h_{\mu\nu}$ are defined through
 the 4D induced metric of Eq.(\ref{INM}). In addition, we need the
 Gauss-Codazzi equations

\beqa
&&\fr{\kappa^2}{2} \left[(K_+)_{\mu\nu}+(K_-)_{\mu\nu}
\right] T^{\mu\nu}=\triangle G_{\mu\nu} n^\mu n^\nu,
 \label{GCE1} \\
&&\kappa^2 {h_{\mu}}^{\lambda} \nabla_{\nu}{T_{\lambda}}^{\nu}={h_{\mu}}^{\lambda}
\triangle G_{\lambda}^{\nu} n_{\nu},
\label{GCE2}
\eeqa
where the last one is nothing but the conservation law
\beq
{d \over d\tau} \left( \varrho a^3 \right)+p{d \over d\tau} \left( a^3
\right)=0.
\eeq
On the other hand, Eq.(\ref{GCE1}) denotes the average of the
values of extrinsic curvature on the two sides. In
Eqs.(\ref{GCE1}) and (\ref{GCE2}), the right-hand sides are
zero because we choose the same cosmological constant for the two
sides.
From Eqs.(\ref{4DI}), (\ref{MAT}) and (\ref{GCE1}), one finds

\beqa
&&(h_{+} \dot t_{+})^{-1}(\ddot a +h'_{+} /2)+
(h_{-} \dot t_{-})^{-1}(\ddot a +h'_{-} /2)=
-\kappa^2 \left (p+\fr{2}{3} \varrho \right)+\fr{b\ell}{2}
\left( -2\fr{\ddot a}{a} +\fr{{\dot a}^2}{a^2}+\fr{k}{a^2}
\right),\label{TPC} \\
&&h_{+}\dot t_{+}+h_{-}\dot t_{-}=\fr{\kappa^2}{3}\varrho a-\fr{b\ell}{2a}
\left( \dot a^2+ k \right),\label{SPC} \\
&&(h_{+} \dot t_{+})^{-1}(\ddot a +h'_{R} /2)-
(h_{-} \dot t_{-})^{-1}(\ddot a +h'_{-} /2)=
3\fr{p}{\varrho a} \left (h_{+}\dot t_{+}- h_{-}\dot t_{-}
\right).
\eeqa
Equation (\ref{TPC}) corresponds to the acceleration part of the
moving domain wall, while Eq.(\ref{SPC}) corresponds to the
velocity of the moving domain wall.
In the case of $b=0$, we can easily solve the above equations
simultaneously\cite{IDA}. However, it is very hard to solve the above
equations as they stand. It is very interesting to derive the Friedmann-like
equation.
For this purpose, we use only  the space component of the junction condition Eq.(\ref{SPC})
which can be rewritten as
\beq
\sqrt{h_+ + \dot a^2}+\sqrt{h_- +\dot a^2}=\fr{\kappa^2}{3}\varrho a-\fr{b\ell}{2a}
\left( \dot a^{2} + k \right).
\label{SEE}
\eeq
Let us introduce $X=\dot a^2 +k$ to derive the Friedmann-like equation from Eq.(\ref{SEE}).
Then this  leads to the fourth-order equation as
\beq
AX^4+BX^3+CX^2+ DX +E=0
\label{4OE}
\eeq
whose coefficients are given by
\beqa
&&A={1\over64}{b^4 \ell^4 \over a^6},\cr
&&B=-{b^2 \ell^2 \over 4a^2}\left( 1+{1\over6}b\kappa^2\varrho\ell
\right),\cr
&&C=\fr{b^2\ell^2}{4 a^2}
\left( -{a^2 \over \ell^2} +\fr{\alpha_{+}+\alpha_{-}}{2a^2}
+{1\over6}\kappa^4\varrho^2 a^2 \right)
+{1\over3}b\kappa^2\varrho\ell, \cr
&&D=-{1\over9}\kappa^4\varrho^2 a^2 +{1\over3}b\kappa^2\varrho\ell
\left( {a^2 \over \ell^2} -\fr{\alpha_{+}+\alpha_{-}}{2a^2}
-{1\over18}\kappa^4\varrho^2 a^2 \right), \cr
&&E=\fr{(\alpha_{+}-\alpha_{-})^2}{4a^2}+\fr{\alpha_{+}+\alpha_{-}}{18}
\kappa^4 \varrho^2+{1\over324}\kappa^8\varrho^4 a^4-
{1\over9}\fr{\kappa^4 \varrho^2 a^4}{\ell^2}.
\eeqa

In the case of $b=0$, the above equation reduces to the
first-order equation and its solution is
\beq
X_0= \dot a^2 +k = -{a^2 \over \ell^2} +\fr{\alpha_{+}+\alpha_{-}}{2a^2}
+{1\over36} \kappa^4 \varrho^2 a^2 +{9\over4}
\fr{(\alpha_{+}-\alpha_{-})^2}{\kappa^4 \varrho^2 a^6}.
\label{ZER}
\eeq
This is the well-known equation for the brane cosmology,
 which has derived in refs.\cite{KRA,IDA,CH,BDL2}. In the case of
$b\not=0$,
we have to solve the fourth-order equation. Although its general
solution is known, it is a formidable task to extract a meaningful
Friedmann-like equation. Here we wish to treat all $b$-terms as
the perturbations around the background (brane cosmology) equation  of
Eq.(\ref{ZER}). In other words, $b$ is considered as a small parameter.
For simplicity we keep the first-order
$b$-terms by considering
$A\to 0$, $B\to 0$, $C\to {1\over3}b\kappa^2\varrho\ell$ as
\beq
{1\over3}b\kappa^2\varrho\ell X^2 +\left(-{1\over9}\kappa^4\varrho^2 a^2
+b\tilde\varrho\right)X +E=0
\eeq
with
\beq
\tilde\varrho=\varrho\left( {a^2 \over \ell^2} -\fr{\alpha_{+}+\alpha_{-}}{2a^2}
+{1\over18}\kappa^4\varrho^2 a^2 \right).
\eeq

Let us propose the perturbed solution as $X=X_0 + b \beta$, then $\beta$
is determined by
\beq
\beta=\fr{3\kappa^2\varrho\ell +9\tilde\varrho}{\kappa^4\varrho^2
a^2}X_{0}.
\eeq
If we express it in terms of the conventional form
\beq
{1 \over2}\dot a^2 +V(a)=-{1 \over2}k
\label{FRE}
\eeq
the potential is given by
\beqa
V(a)= && b\left(\fr{3}{2\kappa^4 \varrho^2\ell^2}-\fr{\kappa^2 \varrho\ell}{24}\right) \cr
&&+ \left[\left(\fr{1}{2}-\fr{\kappa^4\varrho^2\ell^2}{72}\right)
 + b\left( \fr{3}{2\kappa^2\varrho\ell}-\fr{\kappa^2\varrho\ell}{8}
+\fr{\kappa^6\varrho^3 \ell^3}{432}\right)\right]
\fr{a^2}{\ell^2}\cr
&& +\left[
-\fr{1}{2} +b\left( -\fr{3}{2\kappa^2\varrho\ell}+
\fr{\kappa^2\varrho\ell}{18}
\right)\right]\fr{\alpha_{+}+\alpha_{-}}{a^2}\cr
&&-b\fr{3}{4}\fr{(\alpha_{+}+\alpha_{-})^2}{\kappa^2\varrho\ell}{\ell^2\over a^4}\cr
&& +\left[-\fr{9}{8}\fr{(\alpha_{+}-\alpha_{-})^2}{\kappa^4\varrho^2\ell^2}
+ b\left( -\fr{27}{8}\fr{(\alpha_{+}-\alpha_{-})^2}{\kappa^6\varrho^3\ell^3}+
\fr{1}{8}\fr{(\alpha_{+}+\alpha_{-})^2}{\kappa^2\varrho\ell}+
\fr{3}{16}\fr{(\alpha_{+}-\alpha_{-})^2}{\kappa^2\varrho\ell}\right)\right]
{\ell^2\over a^6}\cr
&&-b\fr{27}{8}\fr{(\alpha_{+}-\alpha_{-})^2}{\kappa^6 \varrho^3
\ell^3}{\ell^4\over a^8}\cr
&&+b\fr{27}{16}\fr{(\alpha_{+}-\alpha_{-})^2 \left(\alpha_{+}+\alpha_{-}\right)}
{\kappa^6 \varrho^3 \ell^3}
{ \ell^4\over a^{10}},
\label{POT}
\eeqa
where $\kappa^2\varrho\ell$ is a dimensionless quantity.
For an extremal wall with a fine-tuned tension, we have $\kappa^2\varrho\ell=\kappa^2
\sigma\ell={6\over\ell}\ell=6$. Then its potential reduces to

\beqa
V^{EXT}(a)=&&\left( -{1\over2}+b{1\over12}\right)\fr{\alpha_{+}+\alpha_{-}}{a^2}
-b{(\alpha_{+}+\alpha_{-})^2\over12} \fr{\ell^2}{a^4}\cr
&&\left[-\fr{(\alpha_{+}+\alpha_{-})^2}{32}+b\left(\fr{(\alpha_{+}-\alpha_{-})^2}{64}
+\fr{(\alpha_{+}+\alpha_{-})^2}{48}\right)
\right]\fr{\ell^2}{a^6}\cr
&&-b\fr{1}{8}\fr{\ell^4}{a^8}+
b\fr{(\alpha_{+}-\alpha_{-})^2 \left(\alpha_{+}+\alpha_{-}\right)}
{128}\fr{\ell^4}{a^{10}}.
\eeqa
In the case of $b=0$, we recover the motion of extremal wall with a
fine-tuned tension
Eq.(\ref{FRE}) with\cite{KRA}

\beq
V^{EXT}_{b=0}(a)=-\fr{\alpha_{+}+\alpha{-}}{2a^{2}}-\fr{(\alpha_{+}-\alpha_{-})^{2} \ell^2}
{32 a^6}.
\eeq
The RS static configuration\cite{RS2} can easily be recovered from Eq.(\ref{FRE})
when $k=\alpha_+=\alpha_-=0$.

By the similar way, we can obtain higher-order equation including $b^2,b^3$
and $b^4$ for small $b$. Although we do not derive the Friedmann-like  equation for
 the finite $b$ case (for example, $b=1$ for a counter term),
  our perturbed equation (\ref{FRE}) with (\ref{POT}) is
 valuable for investigating the role of the brane curvature scalar
 in the brane cosmology. This equation corresponds to corrections to the known
 brane cosmology of Eq.(\ref{ZER}).

We observe from Eq.(\ref{POT}) that the non-zero curvature scalar
on the brane induces some interesting consequences. In the first line,
we find new constant terms $b/\varrho, b/\varrho^2$
 which are never found in any (brane) cosmology.
In the second line,  $\varrho^2 a^2$ corresponds to the famous term for
the brane cosmology, but we have additional terms of
$b(1/\varrho, \varrho, \varrho^3)a^2$.
Here we find a standard term of $\varrho a^2$ in the Friedmann
equation. This term never appears if one includes a brane
curvature scalar. The third line represents mass-like terms of $\alpha/a^2$,
 which are derived from  the the electric part of the
Weyl tensor. Here one obtains  new $b$-dependent mass-like terms.
The fourth line is totally a new term. We cannot find  any term of $\alpha^2/a^4$
in the existing literature.
In the fifth one,  we point out that
$\alpha_{+}=\alpha_{-}$ induces a new term such as  $b\alpha^2/a^6$
which never appears in the case of $b=0,Z_2$-symmetric evolution.
In the case of  $\alpha_{+} \not=\alpha_{-}$,
 we get the last two terms of Eq.(\ref{POT}) which are new higher-order terms for $a$
like $b \alpha^2/ a^8, b \alpha^3/a^{10}$ as well as for $\varrho$ like $b/ \varrho^3$.
These may enhance the non-$Z_2$ symmetric evolution.

Finally let us compare our results with the existing literature.
Collins and Holdom in \cite{CH} have derived cosmological
equations for a vacuum bubble expanding into a AdSS$_5$ bulk space
and the edge of a single AdS$_5$ space.  This calculation has been
done without any condition for $b$. As was explained before,
Eq.(\ref{4OE}) is a quartic polynomial in $k + \dot a^2$ which
becomes manageable only for special cases. Hence they studied  two :
 a vacuum bubble and an
edge of a single AdS$_5$ apace. But in the strict sense,
these two cases do not belong to the
brane cosmology.
 Even for an edge universe, they failed to derive a realistic
cosmological evolution for a counter term of  $\sigma=\fr{6}{8 \pi G \ell}, b=1$ because they
found a complex potential.
On the other hand, Deffayet \cite{DEF} investigated  a role of the
brane curvature scalar in Minkowski bulk space. He considered only
the Z$_2$-symmetric evolution within the BDL approach. In this case
he found also the right hand sides of Eqs.(\ref{TPC}) and (\ref{SPC}).
But he took a view that the contribution from the intrinsic
curvature is regarded  as a cosmic fluid of density $\rho_{curv}$ and
pressure $P_{curv}$. In addition he recovered a standard
cosmology in a certain limit. However this picture is slightly different from a
genuine brane cosmology. Although our result has a limitation
such that it is valid only for small $b$, it provides a new
insight to see the role of brane scalar curvature in the brane
cosmology.

We conclude that our equation (\ref{FRE})
corresponds to corrections the known brane cosmology Eq.(\ref{ZER}).
This implies that
brane curvature scalar induces new interesting implications for
the brane cosmology.
\section*{Acknowledgments}

This work was supported in part by the Brain Korea 21 Program, Ministry of
Education, Project No. D-0025 and KOSEF, Project No. 2000-1-11200-001-3.

\end{document}